\documentclass[11pt,a4paper]{article}
\usepackage{amsmath,amssymb,amsfonts}
\usepackage{graphicx,a4wide}
\usepackage[top=3.cm,bottom=3.cm,left=1.5cm,right=1.5cm]{geometry}
\newcommand{\be}{\begin{equation}}
\newcommand{\ee}{\end{equation}}
\newcommand{\bea}{\begin{eqnarray}}
\newcommand{\eea}{\end{eqnarray}}
\newcommand{\nn}{\nonumber}

\newcommand{\Gm}{\Gamma}
\newcommand{\ep}{\epsilon}
\newcommand{\de}{\delta}

\newcommand{\om}{\omega}

\newcommand{\lm}{\lambda}

\newcommand{\sg}{\sigma}

\newcommand{\ov}{\overline}

\newcommand{\oq}{\overline{q}^{\,2}}

\newcommand{\op}{\overline\Pi}

\newcommand{\wt}{\widetilde}
   
\newcommand{\omp}{\om_\pi}
\newcommand{\omh}{\om_h}

\newcommand{\vk}{\vec k}

\newcommand{\vq}{\vec q}

\newcommand{\la}{\langle}
\newcommand{\ra}{\rangle}

\newcommand{\mn}{\mu\nu}

\begin{document}

\title{Observing many body effects on lepton pair production from
low mass enhancement and flow at RHIC and LHC energies}
\author{Sabyasachi Ghosh, Sourav Sarkar and Jan-e Alam}
\maketitle

\begin{center}
\it{Theoretical Physics Division, Variable Energy Cyclotron Centre,\\
1/AF, Bidhannagar, 
Kolkata 700064, India}
\end{center}

\begin{abstract}
The $\rho$ spectral function at finite temperature
calculated using the real-time formalism of thermal field theory
is used to evaluate the low mass dilepton spectra. The analytic structure of
the $\rho$ propagator is studied and contributions to the dilepton
yield in the region below the bare $\rho$ peak from the different
cuts in the spectral function are discussed. The space-time integrated 
yield shows significant enhancement in the region below the bare $\rho$
peak in the invariant mass spectra. It is argued that the variation
of the inverse slope of the transverse mass ($M_T$) distribution can be used 
as an efficient tool to predict the presence of two different
phases of the matter during the evolution of the system. 
Sensitivity of the effective temperature obtained from the slopes of the 
$M_T$ spectra to the medium effects are studied. 
\end{abstract}


\section{Introduction}

The latest results from relativistic $Au+Au$ collisions at Relativistic 
Heavy Ion Collider (RHIC)
have indicated that the matter created in the initial stages in relativistic
heavy ion collisions (HIC) might be in the form
of quark gluon plasma~\cite{npa2005}. This is supported 
by the observation of high transverse momentum ($p_T$) hadron suppression
(jet-quenching) in central collisions compared to the
binary-scaled hadron-hadron interactions~\cite{jetquen}.
The observation of large elliptic flow have also indicated the possibility
of rapid thermalisation of such high density matter~\cite{v2}.

Although the small production cross-section of lepton pairs 
leads to a lower yield, the electromagnetic probes are well suited 
to probe the local properties of the transient form of matter produced
in nuclear collisions at ultra-relativistic energies as they leave the 
system almost unscathed. 
They emanate at all stages and thus are expected to map the 
temperature profile of the evolution. Because of the backgrounds coming from
different stages and the low yield, extraction of signal from the
background becomes a daunting task~\cite{na60}.
Theoretically, the prediction of the yield depends on the
evaluation of the production rate as well as the scenario of space-time
evolution that one employs. The rates of production from the QGP is
controlled by QCD but those from hadronic matter depends on the hadronic 
interactions
that one considers. In the low mass region, the rate of dilepton production 
is controlled by the spectral functions of the vector mesons, specially the
$\rho$ and hence the modification of the $\rho$ spectral function determines
the yield of lepton pairs in this region of invariant mass.

A number of authors have analysed the dilepton spectra from heavy ion 
collisions; the treatments differing both in the construction of the
$\rho$ spectral function as well as the space time evolution scenario
employed. This includes the nature of phase transition,
the equation of state  as well as numerical values of the 
parameters like the initial temperature, the thermalisation time, the
phase transition temperature as well as the chemical and kinetic freeze-out
temperatures. We do not attempt to review or summarise the considerable
amount of work which has been done on this topic except to mention the most
recent few. The NA60 experiment at the CERN SPS measured dimuon pairs in
In-In collisions in which an excess was observed over the contribution 
from hadronic decays at freeze-out in the mass region below the $\rho$ 
peak~\cite{na60}. This was attributed to the broadening of the $\rho$ 
in hot and dense medium~\cite{RappAdv}, in contrast to the earlier
data from the CERES collaboration~\cite{ceres} which is unable
to differentiate between the broadening and the pole shift of the $\rho$
spectral function~\cite{sah}. The NA60 data for the
entire (measured) invariant mass range  is reproduced
by taking into account dilepton productions  
from  Drell-Yan processes,
$q\bar{q}$ annihilation, thermally broadened in-medium
$\rho$, decays of $\rho$ at the freeze-out
surface and primordial $\rho$ produced from the initial hard scattering
~\cite{RappSPS}. 
The dependence of thermal emission on transition and chemical freeze-out
temperatures is also highlighted there. 
The dilepton yield evaluated with the in-medium spectral functions
of $\rho$ and $\omega$ mesons deduced from empirical forward 
scattering amplitudes~\cite{ruppert} does not reproduce the data well
at the low invariant mass ($M<0.5$ GeV) region.
The PHENIX experiment 
reported a substantial excess of electron pairs in the same region of
invariant mass~\cite{phenixdil}. The data has been investigated by several
groups e.g.~\cite{Dusling,Cassing}. The yield in all these cases have remained
insufficient to explain the PHENIX data~\cite{DreesQM}. Thus the issue of low mass 
lepton pair yield still remains an unsettled issue.


In the literature the modification of the $\rho$ propagator comes from
(a) the thermal modification of the decay width into pion pairs and (b)
collisional broadening due to scattering with the excitations in the thermal 
medium~\cite{RappAdv}. In this work we have studied dilepton production 
at RHIC and LHC (Large Hadron Collider) 
energies using a spectral function evaluated in the real time formulation of
thermal field theory with the interaction vertices taken from 
chiral perturbation theory which is the low energy effective theory of QCD.
From the discontinuities of the $\rho$ self energy
associated with the branch cuts in the complex energy plane we provide a
unified description of the apparently different scattering and decay processes
in the medium~\cite{GhoshEPJC} (for a different approach see~\cite{rappprc60,RappAdv}).
The resulting spectral function of the
$\rho$ at non-zero three-momentum shows significant broadening with no appreciable
change in the pole position. Since the dilepton spectra is proportional to the
vector meson spectral function, the $\rho$ in particular, we will attempt
to bring out the medium modifications through different 
aspects of the dilepton spectra (see~\cite{RappAdv,annals} for review).

For the space-time evolution we have used ideal relativistic
hydrodynamics. There are quite a few parameters which go as inputs into this
scheme e.g. the thermalisation time or the initial temperature.
Though studies on elliptic flow of matter produced in such collisions
indicate a rapid thermalisation~\cite{Pasi}, analyses using second order 
transport coefficients with conformal
symmetry indicate 
sizable uncertainties~\cite{Luzum} in its determination. Then comes the issue
of transition temperature in addition to the chemical and kinetic 
freeze-out temperatures. The equation of state, in particular the
velocity of sound is another vital input
which goes into the hydrodynamics and control the evolution profile.
Since electromagnetic probes are emitted throughout the spatial and 
temporal extent of the fireball, the spectra is in fact sensitive
to each of these factors. We will investigate the dependence of
the dilepton spectra in the low mass region on the equation of state (EoS)
obtained from two different schemes.

It is well known that the average magnitude of
radial flow at the freeze-out surface can be extracted from
the $p_T$ spectra of various hadrons.
However, hadrons being strongly interacting objects
can bring the information of the state of the system
when it is too dilute to support collectivity {\it i.e.}
the parameters of collectivity extracted from the hadronic
spectra are limited to the evolution stage where the
collectivity ceases to exist. These collective parameters have
hardly any information about the interior of the matter.
On the other hand dileptons (and real photons)
are produced and emitted from each space time point.
Again, dileptons have an additional advantage because in this
case we have two kinematic variables - of these, the
$p_T$ spectra is affected by the flow in the system
but the $p_T$ integrated invariant mass ($M$) spectra
remains unaltered.
Therefore, if one can identify a domain of $M$ where
QGP contributions dominate then the flow from the interior
of the QGP phase may be extracted by studying the 
$M_T$ spectra at this $M$ window {\it i.e.}
a judicious choice of $p_T$ and
$M$ windows will be very useful to characterise the
flow in QGP and hadronic phases, which will shed light on
the time evolution of the collectivity in the system~\cite{Deng}.
The effective slope of the dilepton spectra have been extracted both
for the RHIC and LHC energies to demonstrate these aspects.

In this work we have thus focused on two kinds of many body 
effects which
are inherent in the evaluation of dilepton spectra from heavy ion
collisions. The one concerning the in-medium vector meson spectral function
is microscopic in nature originating from the dynamics of effective 
hadronic interactions taken from chiral perturbation theory 
and the other is a macroscopic manifestation which results in
collective flow of the fireball described by relativistic hydrodynamics.
Both of these effects can be observed by studying the 
low mass lepton pairs with appropriate choice of 
invariant mass and transverse momentum windows. We demonstrate
this by evaluating the lepton pair spectra for RHIC and LHC conditions. 

The organisation of the paper is as follows. In section 2
we shall discuss the details of calculation of the spectral
function of vector mesons
and subsequently the differential dilepton rate.
Space-time evolution of the evolving matter will be
discussed in section 3.
Space-time integrated dilepton yields
are presented in section 4 and finally we conclude in section 5.

\section{Dilepton emission rate}

The rate of dilepton production rate from a thermal medium produced in heavy 
ion collisions is well known to be given by~\cite{Larry}
\be
\frac{dN}{d^4qd^4x}=-\frac{\alpha^2}{6\pi^3q^2}
L\left(M^2\right)f_{BE}(q_0) g^{\mn}\,
W_{\mn}\left(q_0,\vq \right)
\label{eq:defrate}
\ee
where the factor $L(M^2)=(1+{2m_l^2}/{M^2})~
(1-4m_l^2/M^2)^{1/2}~$ is of the order of unity for electrons,
$M(=\sqrt{q^2})$ being the invariant
mass of the pair and the electromagnetic (e.m.) current correlator
$W_{\mn}$ is defined by
\be
W_{\mn}(q_0,\vq)=\int d^4x\,e^{iq\cdot x}\la [J^{em}_\mu(x),J^{em}_\nu(0)]\ra
\label{eq:wmn}
\ee
Here $J^{em}_\mu(x)$ is the electromagnetic current and $\la\ra$ indicates ensemble
average.
For a deconfined thermal medium such as the QGP, Eq.~(\ref{eq:defrate})
leads to the 
standard rate for lepton pair production from  
$q\bar q$ annihilation~\cite{cleymans} at lowest order. 

At low values of the invariant mass $M$, the
electromagnetic current of quarks
$J^{em}_\mu=\frac{2}{3}\bar u\gamma_\mu u-\frac{1}{3}\bar d\gamma_\mu 
d-\frac{1}{3}\bar s\gamma_\mu s$ may be decomposed as
\be
J^{em}_\mu=J^\rho_\mu+J^\omega_\mu/3-J^\phi_\mu/3 
\ee
where the vector currents 
\bea
J^\rho_\mu &=&\frac{1}{2}(\bar u\gamma_\mu u -\bar d\gamma_\mu d)\nonumber\\
J^\omega_\mu &=&\frac{1}{2}(\bar u\gamma_\mu u +\bar d\gamma_\mu d)\nonumber\\
J^\phi_\mu&=&\bar s \gamma_\mu s
\label{eq:jv}
\eea
are named by the lowest mass hadrons $\rho^0,\ \omega$ and
$\phi$ in the corresponding channel.
Defining the correlator of these currents $W_{\mn}^{\rho,\om,\phi}$ analogously
as in (\ref{eq:wmn}) we write,
\be
W_{\mn}=W^\rho_{\mn}+W^\omega_{\mn}/9+W^\phi_{\mn}/9~.
\ee
One now has to specify the coupling of the
currents to the corresponding vector fields. For this
purpose we write, in the narrow width approximation~\cite{Shuryak},
\be
\la 0|J^{em}_\mu(0)|R\ra=F_R m_R \ep_\mu
\label{eq:shuryak}
\ee
where $R$ denotes the resonance in a particular channel and $\ep_\mu$ is the
corresponding polarisation vector.
The coupling constants $F_R$ are obtained 
from the partial decay widths into $e^+e^-$ through the relation
\be
F_R^2=\frac{3m_R\Gamma_{R\to e^+e^-}}{4\pi\alpha^2}
\ee
yielding $F_R$=0.156 GeV, 0.046 GeV and 0.079 GeV for $\rho$, $\omega$ and
$\phi$ respectively.
Eq.~(\ref{eq:shuryak}) suggests the operator
relations
\be
J_\mu^{\rho}(x)=F_\rho m_\rho V_\mu^{\rho}(x),~~J_\mu^{\om}(x)=3F_\om m_\om V_\mu^{\om}(x)
~~{\rm etc.}
\ee
where $V_\mu^{\rho(\om)}(x)$ denotes the field operator for the $\rho(\om)$ meson.
The vector correlator in the $\rho$ channel for example, is then given by
\bea
W^\rho_{\mn}&=&F_\rho^2 m_\rho^2\int d^4x\,e^{iq\cdot x}\la
 [V_\mu^\rho(x),V_\nu^\rho(0)]\ra\nonumber\\
&=&K_\rho\, \rho^\rho_{\mn}(q_0,\vq)
\label{eq:wrho}
\eea
where $\rho^\rho_{\mn}$ is the spectral function of the $\rho$ meson
in the thermal medium and $K_\rho=F_\rho^2 m_\rho^2$.
Defining the spectral functions of $\om$ and $\phi$ as in ($\ref{eq:wrho}$), 
the dilepton rate is given by
\be
\frac{dN}{d^4qd^4x}=-\frac{\alpha^2}{6\pi^3q^2}
f_{BE}(q_0) g^{\mn}
[K_\rho \rho^\rho_{\mn}(q_0,\vq)+K_\om \rho^\om_{\mn}(q_0,\vq)
+K_\phi \rho^\phi_{\mn}(q_0,\vq)]
\label{eq:dilrate2}
\ee 

\subsection{Spectral function in terms of self energy}

\begin{figure}
\centerline{\includegraphics[scale=0.65]{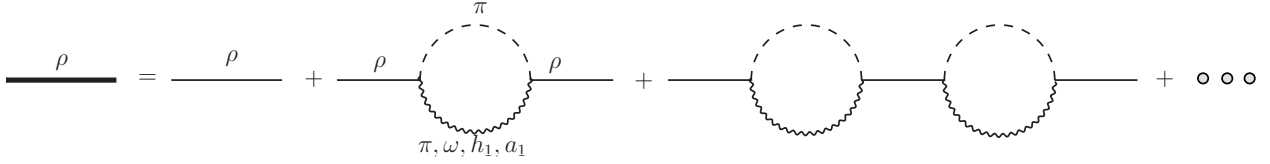}}
\caption{The Dyson equation for the $\rho$ propagator, the small circles
indicating terms of higher order in the series.}
\label{fig:dyson}
\end{figure}

The $\rho$ meson plays the most significant role in the low mass region.
We thus begin with the complete $\rho$ propagator which in the
real time formulation of thermal field theory is a $2\times 2$ matrix
defined by
\be
D^{ab}_{\mn}(q_0,\vq)=i\int d^4x\,e^{iq\cdot x}\la
 {\cal {T}}_c V_\mu^\rho(x) V_\nu^\rho(0)\ra^{ab}
\label{eq:dab} 
\ee
where the superscripts $(a,b=1,2)$
are thermal indices and ${\cal{T}}_{c}$ denotes time ordering 
with respect to a contour in the complex time plane. 
In the case of a symmetric contour this matrix function
factorises  as~\cite{Bellac}
\be
D^{ab}_{\mn}=U\left(\begin{array}{cc}\ov D_{\mn} & 0\\0 & -\ov D_{\mn}^*\end{array}
\right)U~;~~~~~~~~~U=\left(\begin{array}{cc}\sqrt{1+n} & \sqrt{n}\\
\sqrt{n} & \sqrt{1+n}\end{array}\right)
\ee
and so is given essentially by a single analytic function, the diagonal
component $\ov D_{\mn}$.
This function admits a Kallen Lehmann representation given by~\cite{MallikRT}
\be
\ov D_{\mn}(q_0,\vq)=\int_{-\infty}^{\infty}\frac{dq_0'}{2\pi}
\frac{\rho_{\mn}^\rho(q_0',\vq)}{q_0'-q_0-i\eta\ep(q_0)}
\ee
in terms of the spectral function defined in Eq.~(\ref{eq:wrho}).
From this we obtain
\be
\rho_{\mn}^\rho(q_0,\vq)=2\ep(q_0){\rm Im}\ov D_{\mn}(q_0,\vq)
\label{eq:rhoImD}
\ee
Thus the dilepton rate is essentially given by the imaginary part of the $\rho$ propagator
which we will now elaborate in the following.

The complete $\rho$ propagator in the medium is obtained from 
the Dyson equation 
\be
D_{\mn}^{ab}(q)=D_{\mn}^{(0)ab}(q)- D_{\mu\lm}^{(0)ac}(q)\Pi^{cd,\lm\sg}(q)
 D_{\sg\nu}^{db}(q)
\label{eq:dyson_Dab}
\ee
pictorially seen in Fig.~\ref{fig:dyson}.
$D_{\mn}^{(0)ab}(q)$ is the free thermal propagator and $\Pi_{\mn}^{ab}(q)$ 
denotes self energy insertions both of which are
$2\times 2$ matrices in this approach. 
Here we have considered
one-loop diagrams consisting of a pion and another hadron $h$ where 
$h$ stands for $\pi$, $\om$, $a_1$ ,$h_1$ mesons
restricting thus to the non-strange meson sector. These self energy graphs
provide corrections to the $\rho$ meson propagation in the medium and will
be evaluated in the next section.

One can get rid of the thermal indices on 
diagonalising the thermal propagator and self energy matrices so that
Eq.~(\ref{eq:dyson_Dab}) now takes the form
\be
\ov D_{\mn}(q)=\ov D_{\mn}^{(0)}(q)- \ov D_{\mu\lm}^{(0)}(q)\ov\Pi^{\lm\sg}(q)
\ov D_{\sg\nu}(q)
\label{eq:dyson_Dbar}
\ee
the bars denoting the diagonal components as discussed above. 
$\ov D_{\mn}^{(0)}(q)$ turns out to be the vacuum propagator given by
\be
\ov D_{\mn}^{(0)}(q)=\left(-g_{\mn}+
\frac{q_\mu q_\nu}{m_\rho^2}\right)\frac{-1}
{q^2-m_\rho^2+i\ep}
\ee

The complete propagator $\ov D_{\mn}(q)$ can be obtained
in terms of transverse and 
longitudinal components by writing
\be
\ov D_{\mn}=P_{\mn}\ov D_t + Q_{\mn}\ov D_l
\ee
and
\be
\op_{\mn}=P_{\mn}\op_t + Q_{\mn}\op_l
\ee
where the projection operators are given by
\be
P_{\mn}=-g_{\mn}+\frac{q_\mu q_\nu}{q^2}-\frac{q^2}{\oq}\wt u_\mu \wt
u_\nu, ~~~~~Q_{\mn}=\frac{(q^2)^2}{\oq}\wt u_\mu \wt
u_\nu,~~~~\oq=(u\cdot q)^2-q^2,
\label{defPQ}
\ee
$u_\mu$ being four-velocity of the heat bath and 
$\wt u_\mu=u_\mu-u\cdot q\,q_\mu/q^2$.
Using these in Eq.~(\ref{eq:dyson_Dbar}) we eventually arrive at the solutions
\be
\ov D_t(q)=\frac{-1}{q^2-m_\rho^2-\op_t(q)},~~~~~
\ov D_l(q)=\frac{1}{q^2}\frac{-1}{q^2-m_\rho^2-q^2\op_l(q)}
\ee
neglecting the non-transverse piece in $\ov D_{\mn}^{(0)}(q)$. 
These are used in Eqs.~(\ref{eq:rhoImD}) and (\ref{eq:dilrate2})
to arrive at the dilepton emission rate
\be
\frac{dN}{d^4qd^4x}=\frac{\alpha^2}{\pi^3q^2}f_{BE}(q_0) \left[
{K_{\rho}} A_\rho(q_0,\vq)
+{K_{\omega}} A_\omega(q_0,\vq)
+{K_{\phi}} A_\phi(q_0,\vq)\right]
\label{eq:dilrate3}.
\ee
where e.g. $A_\rho(=-g^{\mn}\rho^\rho_{\mn}(q_0,\vq)/6)$ is given by
\bea
A_\rho&=&-\frac{1}{3}\left[\frac{2\sum{\rm
Im}\Pi^R_t}{(q^2-m_\rho^2-\sum\mathrm{Re}\Pi^R_t)^2
+(\sum{\rm Im}\Pi^{R}_t)^2}\right.\nonumber\\&&\left.+\frac{q^2\sum{\rm Im}\Pi^R_l}
{(q^2-m_\rho^2-q^2\sum\mathrm{Re}\Pi^R_l)^2
+q^4(\sum{\rm Im}\Pi^{R}_l)^2}\right]
\label{eq:spdef}
\eea
the sum running over the $\pi-h$ loops.
Thus, the dilepton emission rate in the present scenario boils down 
to the evaluation of the self energy graphs shown in Fig.~\ref{fig:dyson}. 
The real and imaginary parts of the self energy function
can be obtained from the 11-component 
as~\cite{Bellac}
\bea
&&{\rm Re}\Pi^R_{\mn}(q_0,\vq)={\rm Re}\ov\Pi_{\mn}(q_0,\vq)={\rm Re}\Pi^{11}_{\mn}(q_0,\vq)\nonumber\\
&&{\rm Im}\Pi^R_{\mn}(q_0,\vq)=\ep(q_0){\rm Im}\ov\Pi(q_0,\vq)=\tanh(\beta q_0/2){\rm Im}
\Pi^{11}_{\mn}(q_0,\vq)
\eea
where $\Pi^R$ denotes the retarded self-energy.

\subsection{Evaluation of $\rho$ self energy}
\vskip 0.1in
\noindent{\bf A. Hot meson gas}
\vskip 0.1in
The 11-component of the thermal self-energy matrix for the $\pi-h$ loops is given by
\be
\Pi_{\mn}^{11}(q)=i\int\frac{d^4k}{(2\pi)^4}N_{\mn}(q,k)D_\pi ^{11}(k)D_h^{11}(q-k)~,
\ee
where $D^{11}(q)$ are the thermal propagators~\cite{MallikRT}
and the factor $N_{\mn}$ 
includes tensor 
structures associated with the two vertices and those in the vector propagator.
In Ref.~\cite{GhoshEPJC}, this was evaluated in detail using interaction
vertices from chiral perturbation theory to obtain  
the imaginary part of the retarded $\rho$ self energy. For positive
values of $q_0$ this is given by
\bea
&&{\rm Im}\Pi_{\rho\pi}(q_0,\vq)=-\pi\int\frac{d^3\vec k}{(2\pi)^3 4\omp\omh}\times\nonumber\\
&&\left[N(k_0=\omp)(1+n(\omp)+n(\omh))\de(q_0-\omp-\omh)
+N(k_0=-\omp)(n(\omp)-n(\omh))\de(q_0+\omp-\omh)\right]~.
\label{eq:impi}
\eea
where $n$ is the Bose distribution
function and the energy variables are
$\omp=\sqrt{m_\pi^2+\vk^2},\,\,\omh=\sqrt{m_h^2+(\vq-\vk)^2}$.
Here we have suppressed indices corresponding to the longitudinal
and transverse components which have been evaluated as in~\cite{GhoshEPJC}.  
We recall that the regions in the $q_0$ plane in which the two terms are non-vanishing
 give rise to cuts in the 
self energy function. These are controlled by the corresponding delta functions.
The first term is non-vanishing for $q^2\ge (m_h+m_{\pi})^2$
producing the unitary cut and
the second term is non-vanishing for $q^2\ge (m_h-m_{\pi})^2$ giving 
the Landau cut.
The unitary cut is also present in vacuum but the 
Landau cut appears only in the medium and arises from scattering of $\rho$
with the particles present there. With the aid of the delta functions
the integration over $\vec k$ is easily performed.  

Note that the heavy mesons in~\cite{GhoshEPJC} have been treated in the narrow width
approximation. In the following we consider the conventional 
prescription (see e.g.
\cite{Nagahiro}) to take into account the decay widths of the $h_1$ and
$a_1$ mesons without disturbing the analytic structure 
discussed above. 
Here, the self energy functions are convoluted with the (vacuum) spectral
functions of the heavy mesons $(h)$ as

\be
\Pi(q,m_h)=\frac{1}{N_h}\int_{(m_h-2\Gm_h)^2}^{(m_h+2\Gm_h)^2}
d\tilde m_h^2\frac{1}{\pi}{\rm Im}\frac{1}{\tilde m_h^2-m_h^2+i\tilde
m_h\Gm(\tilde m_h)}\, \Pi(q,\tilde m_h)
\ee
with
\be
N_h=\int_{(m_h-2\Gm_h)^2}^{(m_h+2\Gm_h)^2}d\tilde m_h^2\frac{1}{\pi}
{\rm Im}\frac{1}{\tilde m_h^2-m_h^2+i\tilde m_h\Gm(\tilde m_h)}
\ee
where $\Gm_h$ are the measured~\cite{PDG} decay widths 
and  
$\Gm(\sqrt{s}=\tilde m_h)$ are the corresponding ones calculated 
using hadronic interactions. As a consequence of this
convolution the clear distinction between the regions of non-zero
imaginary part will be smeared and for sufficiently large
width might even appear continuous. However, for a particular value of
the mass $\tilde m_h$, the imaginary part receives contribution from only one
of the cuts.

\vskip 0.1in
\noindent{\bf B. Effect of baryons}
\vskip 0.1in
It is known that the scattering of the $\rho$ with baryons also
contribute to the broadening of its spectral function even in the
case of zero baryon density, and thereby increase the production of lepton pairs
at invariant masses below the $\rho$ peak. The relevant quantity in this case is
the $\rho$ self-energy due to baryon loops which has been evaluated at finite
temperature and density using well established hadronic interactions in a many
body approach~\cite{RappNPA}. 

The $\rho$ self-energy at finite temperature and baryon density has also been
estimated in terms of empirical scattering amplitudes~\cite{Eletsky} for $\rho$
scattering from nucleons using experimental data.  The self energy of the $\rho$ 
is obtained from these amplitudes using a low density approximation as

\be
\Pi_{\rho N}(E,p)=-4\pi\int\frac{d^3k}{(2\pi)^3}n_{B}(\omega)\frac{\sqrt{s}}{\omega}f^{c.m.}_{\rho
N}(s)
\ee
The scattering amplitude $f^{c.m.}_{\rho N}(s)$ in the low energy region is
described in terms of $N^*$ and $\Delta $  baryon resonances and a Regge model
is employed at higher energies~\cite{Eletsky}. Since the spectral
function in this approach seems to agree with the many body 
approach as shown in~\cite{RappJPG} we will adopt this formalism to estimate the
baryonic contribution to the self energy.

The total $\rho$ self energy in hadronic matter containing mesons and baryons is then written
as
\be
\Pi_\rho=\Pi_{\rho h}+\Pi_{\rho N}~.
\ee

\begin{figure}
\centerline{\includegraphics[scale=0.6]{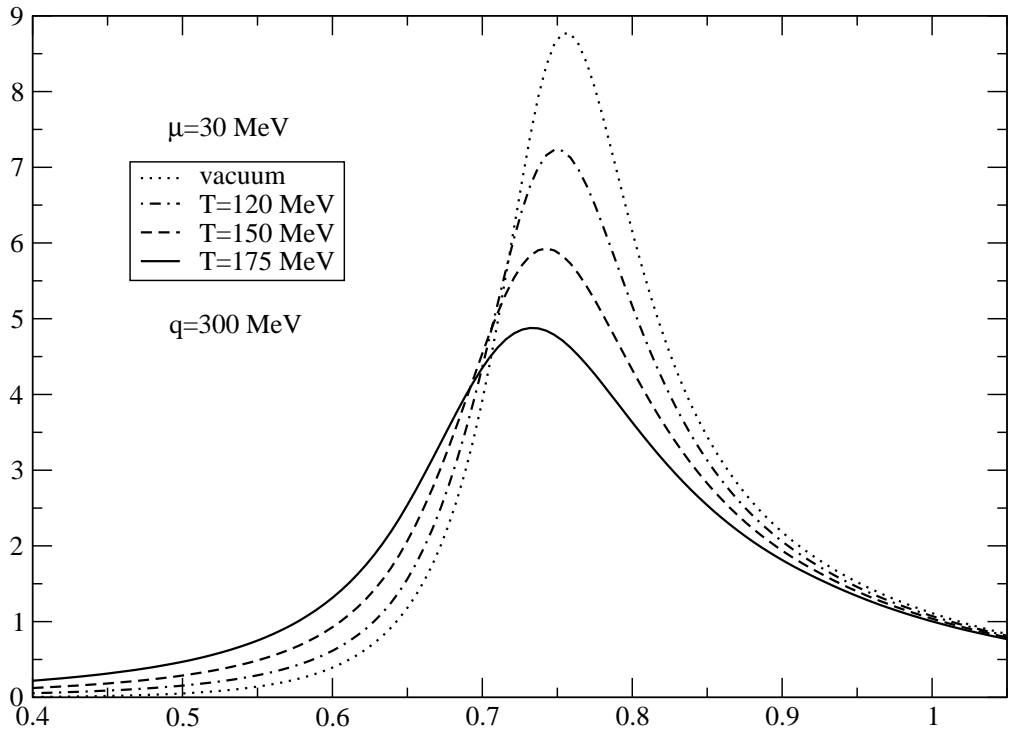}}
\caption{The $\rho$ spectral function averaged over the transverse and
longitudinal polarisations for $\vq$=300 MeV and $\mu_B=30$ MeV.}
\label{fig:spec3d}
\end{figure}

The spectral function as a function of $M$ is plotted
in Fig.~\ref{fig:spec3d} for a range of temperatures at a baryonic chemical potential
$\mu_B=30$ MeV relevant for RHIC energies. 
We observe significant broadening  at higher temperatures with almost negligible shift in the
pole mass.

\subsection{The dilepton emission rate at temperature $T$}

\begin{figure}
\centerline{\includegraphics[scale=0.5]{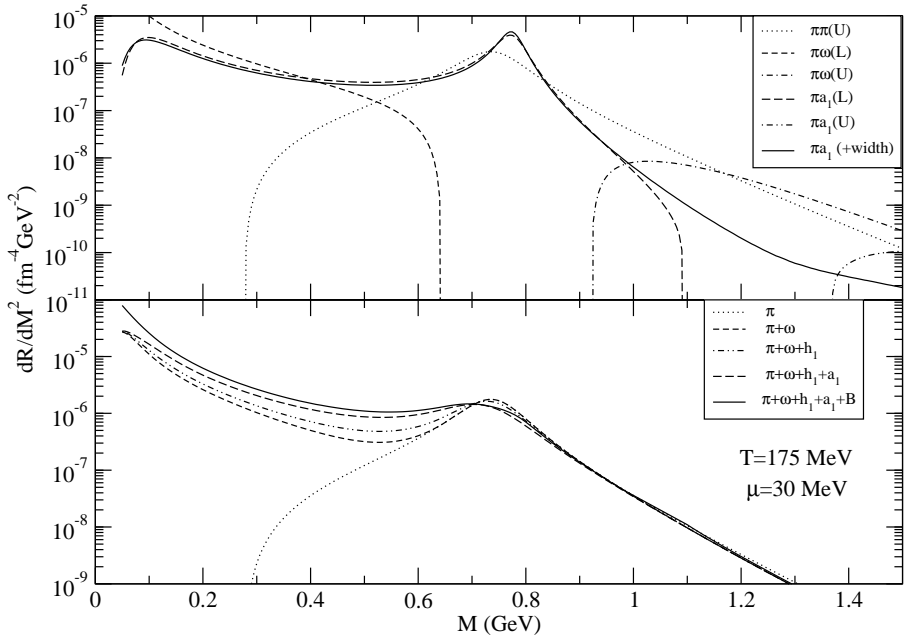}}
\caption{Upper panel shows contributions from the discontinuities of the  self-energy graphs
to the dilepton emission rate at $T=175$ MeV and $\mu_B=30$ MeV. $L$ and $U$ denote the
Landau and unitary cut contribution. Lower panel shows contributions from
the mesons and baryons.}
\label{fig:dil_LU}
\end{figure}

We next plot in Fig.~\ref{fig:dil_LU}, upper panel, the dilepton emission rate keeping only the 
$\rho$ contribution in Eq.~(\ref{eq:dilrate3}) in which we show, for
illustrative purposes, the 
relative contributions from
the cuts in the $\pi-h$ loops keeping only one of them at a time. 
The unitary and Landau cuts for
the $ \pi,\omega,h_1 $ and $a_1$ are seen to contribute with different magnitudes
for different values of the energy and three momenta of the off-shell $\rho$.
In the time-like region, in the vicinity of the (bare)
rho mass the imaginary part of the self energy from a particular loop     
receives dominant contribution from only one of the cuts. The $\pi-\pi$ loop
for example, has only the unitary cut and this contributes most significantly
to dilepton emission near the $\rho$ pole. In contrast, the Landau cut contribution from
the $\pi-\om$ loop is dominant up to about 400 MeV. 
Since this cut ends at $M=m_\om-m_\pi$ and
the unitary cut starts at $M=m_\om+m_\pi$ there is no contribution 
at the $\rho$ pole. 
However, the unitary cut of the $\omega\pi$ loop could make a significant 
contribution as seen in~\cite{GhoshEPJC,vanHeesPRL}.
The Landau cut for the $\pi-a_1$ self-energy extends up to about 1100 MeV and
makes a substantial contribution both at and below the $\rho$ pole. The unitary cut
starts at a much higher value of $M$ and hence does not make a significant contribution
to the $\rho$ spectral function. We also show the effect of convolution 
over the width of the $a_1$ as discussed above. As expected, the contributions
from the Landau and unitary cuts are now joined by a continuous line, the
boundaries being smeared out due to the substantial width of the $a_1$. 
While analysing
the different contributions one must keep in mind that the 
total contribution from the different loops to the spectral function
 is not a linear sum of
the individual contributions as is clear from the definition
 given in Eq.~(\ref{eq:spdef}). This is seen
in the lower panel where 
the cumulative contribution to the lepton pair yield is shown
for the $\pi-\pi$ and $\pi-h$ loops. Also shown is the 
enhancement in yield obtained by including baryons at $\mu_B=$ 30 MeV 
for RHIC energies. 

\begin{figure}
\centerline{\includegraphics[scale=0.5]{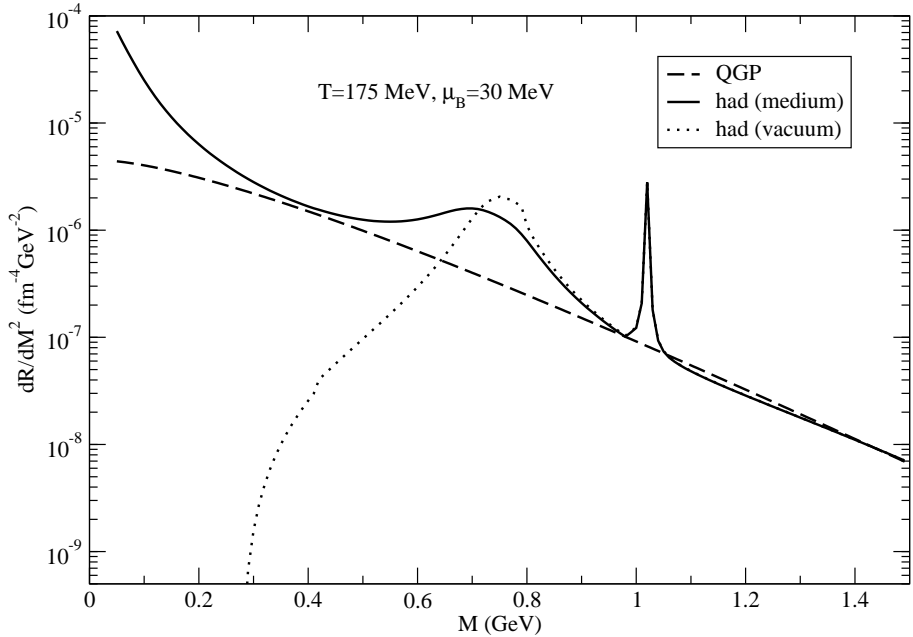}}
\caption{The dilepton emission rate from different sources at $T=175$ MeV and $\mu_B=30$ MeV.}
\label{fig:dilrate}
\end{figure}

We have also included dilepton emission from the $\omega$ and the $\phi$. The 
width of the $\omega$ at finite temperature is taken from the calculation of
Ref~\cite{Weise} where a framework similar to the one employed here has
been used. For the $\phi$ only the vacuum width has been considered. 
In addition, the spectral function of the $\rho$ and $\omega$ 
has been augmented with a continuum contribution given by
\be
\frac{dN}{d^4qd^4x}=\frac{\alpha^2}{\pi^3}f_{\mathrm{BE}}(q_0)
\sum_{V=\rho,\omega}A_V^{\rm cont}
\label{continuum}
\ee
where the continuum part is parametrised as~\cite{annals,Shuryak}
\be
A_\rho^{\rm cont}=\frac{1}{8\pi}\left(1+\frac{\alpha_s}{\pi}\right)
\frac{1}{1+\exp(\omega_0-q_0)/\delta}
\ee
with $\omega_0=1.3$ (1.1) for $\rho$ ($\omega$) GeV and $\delta=0.2$ for for
both $\rho$ and $\omega$. The continuum contribution for the $\omega$ contains
an additional factor of 1/9.
The dilepton emission rates from QGP and hadronic matter have been plotted in 
Fig.~\ref{fig:dilrate} at a temperature of 175 MeV and baryonic chemical potential
of 30 MeV. We observe significant
enhancement in the dilepton yield in the mass region below the $\rho$ pole 
compared to vacuum. This rate has been used in the analysis of the dimuon spectra
obtained from In-In collisions at 17.3 GeV at CERN SPS~\cite{nayak}. The calculations
show a reasonable agreement with the invariant mass spectra for different
$p_T$ ranges as well as the $M_T$ spectra for different $M$ bins.

\section{Space time evolution}

Thus far we have discussed the dilepton emission rates for a given
temperature. In a HIC the temperature corresponds to that of local equilibrium 
and is hence a function of position and time. For a quantitative evaluation
of the dilepton yield one has to convolute the static rate over
space and time resulting in an integrated yield which is a superposition
of contributions from a range of temperatures. 
Ideal relativistic hydrodynamics is 
used to carry out this scheme in which a
number of inputs are required. These are discussed in the following.

\subsection{Equation of state and initial conditions}
The evolution of the fluid is governed by the energy momentum
conservation equation
\be
\partial_\mu\,T^{\mn}=0
\label{eom1}
\ee
where $T^{\mn}=(\epsilon+P)u^{\mu}u^{\nu}\,+\,g^{\mn}P$ is the energy
momentum tensor for ideal fluid.  This 
together with a relation connecting the pressure $P$ and the energy density 
$\epsilon$, known as the EoS provides a closed set
of equations~\cite{hvg}. With cylindrical symmetry and longitudinal
boost invariance~\cite{bjorken}, these set of equations provide the energy density
and transverse velocity as a function of the proper time and the
radial coordinate. The initial energy density and  radial velocity profiles 
which go as inputs are taken as:
\be
\epsilon(\tau_i,r)=\frac{\epsilon_0}{1+e^{(r-R_A)/\delta}}
\label{enerin}
\ee
 and
\be
v(\tau_i,r) = 0
\label{vrin}
\ee
$\delta$ ($\sim 0.5$ fm here) is a parameter, known as the surface thickness. 
The static
rate, $dR/dM^2$ at $M=0.4$ GeV with $\mu_B=30$ MeV and $\mu_B=0$ differs
by less than $10\%$. The difference will be even 
smaller at LHC energies because
of the smaller value of $\mu_B$ at central rapidity region.
Therefore, the effects of $\mu_B$ on space-time evolution is neglected
in the present work.

It is natural to expect that different EoS's will govern the hydrodynamic
flow quite differently~\cite{Pasi} and as far as the search for QGP is concerned, 
the goal is to look for
distinctions in the observables due to the different EoS's (corresponding
to the novel state of QGP vis-a-vis that for the usual hadronic matter). 
It is thus imperative to understand in what respects the two EoS's differ 
and how they affect the evolution in space and time. In order to check
sensitivity with the equation of state (EoS), we have considered
two scenarios: (a) hadronic resonance gas (HRG) with all hadrons up to mass 
2.5 GeV for the hadronic phase
along with a bag model EoS for the QGP phase and (b)
EoS obtained from lattice QCD calculations (LQCD)~\cite{MILC}.

One of the most important parameters that go into the space time
evolution are the values of the initial temperature and the thermalisation
time. There are indications that QGP thermalises quite early at RHIC energies. 
In case of isentropic expansion the experimentally measured hadron 
multiplicity can be related to the initial temperature and thermalisation time 
by the following equation~\cite{hwa}:
\be
T_i^3(b_m)\tau_i=\frac{2\pi^4}{45\zeta(3)\pi\,R_{\perp}^2 4a_k}
\langle\frac{dN}{dy}(b_m)\rangle
\label{eq:intemp}
\ee
where $\langle dN/dy(b_m)\rangle$ is the hadron (predominantly pion) 
multiplicity
for a given centrality class with maximum impact parameter $b_m$. 
$R_{\perp}$ is the transverse dimension of the system,
$\tau_i$ is the initial thermalisation time,
$\zeta(3)$ is the Riemann zeta function and
$a_k=({\pi^2}/{90})\,g_k$ is related to the degeneracy ($g_k$) of 
the system created.
The hadron multiplicity  resulting from $Au + Au$ collisions
is related to that from $pp$ collisions at a given
impact parameter and collision energy by
\be
\langle \frac{dN}{dy}(b_m)\rangle=\left[(1-x)\langle N_{part}(b_m)\rangle/2
+x\langle N_{coll}(b_m)\rangle\right]\frac{dN_{pp}}{dy}
\label{eq:dndy}
\ee
where $x$ is the fraction of hard collisions, $\langle N_{part}\rangle$ and
$\langle N_{coll}\rangle$ are the
average numbers of participants and collisions respectively 
evaluated by using Glauber model.
$dN_{pp}^{ch}/dy= 2.5-0.25ln(s)+0.023ln^2s$,
is the multiplicity of the produced hadrons
in $pp$ collisions at centre of mass energy, $\sqrt{s}$~\cite{KN}.
Assuming $10\%$ hard (i.e. $x=0.10$) and
$90\%$ soft collisions for initial entropy production the value
of $dN_{pp}^{ch}/dy$  turns out to be about 2.43 at $\sqrt{s}=200$ GeV.  
For RHIC energy, we take $T_i=320$ MeV with initial time $\tau_i=0.2$ fm/c
which acts as inputs to the hydrodynamic evolution.

At LHC the measured values of $dN^{ch}_{pp}/dy$
for $\sqrt{s_{\mathrm NN}}=900$ GeV, 2.36 TeV and 7 TeV 
are 3.02, 3.77 and 6.01 respectively~\cite{alice}. 
The value  $dN^{ch}_{pp}/dy$ at $\sqrt{s_{\mathrm NN}}=5.25$ TeV 
is obtained by interpolating the above experimental data mentioned above.
Assuming $x=0.2$ in Eq.~(\ref{eq:dndy}) we obtain $dN/dy=2607$ in Pb+Pb 
collision for 0-10\% centrality. For $\tau_i=0.1$ fm/c we get $T_i=686$ MeV.

For the space-time picture, we thus work in the following
scenario. An equilibrated
QGP is formed at initial temperature (time) $T_i (\tau_i)$, the system then
cools due to expansion and when the temperature reaches $T_c$ it
undergoes a phase transition from QGP to hadrons.
After the completion
of the phase transition the hadronic matter cools and eventually freezes out 
first chemically  at a temperature $T_{ch}$ and then kinetically
at a temperature $T_F$.  
The transition temperature is taken as $ T_c\sim $ 175 MeV.
The other inputs which goes into the calculations are 
chemical ($T_{\mathrm ch}$)
and kinetic freeze-out ($T_F$) temperatures. The kinetic freeze-out 
in the system occurs when both the elastic and in-elastic 
collisions stop {\i.e.} the freeze-out takes place when the 
collectivity in the system ceases to exist. 
The value of  $T_F$ can be constrained from the hadronic $p_T$ spectra
~\cite{npa2005}. In the present work we take $T_F = 120$ MeV
which reproduces the $p_T$ spectra of pions, kaons
reasonably well~\cite{jpgnpa}.
The ratios of various hadrons measured experimentally at
different $\sqrt{s_{\mathrm {NN}}}$
indicate that the system formed in heavy ion collisions chemically decouple
at $T_{\mathrm {ch}}$ which is higher than $T_F$~\cite{pbm}.  
Therefore, the system remains out of chemical equilibrium
from $T_{\mathrm {ch}}$ to $T_F$. The deviation of the system from
the chemical equilibrium is taken in to account by introducing
chemical potential for each hadronic species~\cite{bebie}.
The chemical non-equilibration affects the yields
through the phase space  factors of the hadrons which in turn
affects the productions of the EM probes. The chemical
potential, $\mu_j$  for the  hadronic species $j$ as 
a function of $T$ have been taken from Ref.~\cite{hirano}:
\begin{equation}
\frac{n_j(T,\mu_j)}{s(T,\{\mu_j\})}=
\frac{n_j(T_{ch},\mu_j=0)}{s(T_{ch},\{\mu_j\}=0)}
\end{equation}
where $n_j$ is the density of hadron $j$ contains direct 
as well as contributions from resonance decays.  
The $\mu_j$ is a function of $T$ and it vanishes at $T=T_{ch}$ 
($=170$ MeV here). 
Therefore, the space time evolution of $\mu_j$ is
dictated by the evolution of $T$. The chemical potentials of pions,
$\omega$, $h_1$, $a_1$, $\phi$ and proton enters through their
thermal distributions as a fugacity factor. The values of
the respective chemical potential at the kinetic freeze-out temperature,
$T_F=120$ MeV are $\mu_\pi=68$ MeV, $\mu_\omega=179$ MeV, $\mu_{h_1}=204$ MeV
$\mu_{a_1}=204$ MeV, $\mu_\phi=252$ MeV  $\mu_{\mathrm proton}=258$ MeV.
  
We have also included the contribution to
the dilepton yield from the decays of $\rho$ mesons at freeze-out using
the Cooper-Frye formula~\cite{cooper} as follows.
For a special case of unstable vector mesons  
we need to know the thermal phase space factor 
corresponding to  an unstable Boson which is given by 
\begin{equation}
f_{\mathrm {unstable}}=
\frac{g}{(2\pi)^3}\,\int\frac{1}{\exp(E/T) - 1}A_\rho(M)dM^2
\end{equation}
where $E=\sqrt{p^2+M^2}$, $g$ is the statistical degeneracy and 
$A_\rho(M)$ is the (vacuum) spectral function of the vector meson under 
consideration. For stable particle $A_\rho(M)$ reduces to a Dirac
delta function and consequently the usual phase space factor for
a stable particle is  recovered upon integration over $dM^2$. 
Therefore, the $m_T$ distribution of dimuons from vector meson decay after the
freeze-out is given by~\cite{RappSPS} (see also~\cite{dusling09}):
\bea
\frac{dN_{\gamma^\ast}}{M_{T}dM_{T}dM^2dy}&=&2\pi\int dr \int d\eta \int d\phi\, 
r\tau\nn\\
&&\times\left(M_T\mathrm {cosh}(y-\eta)-\frac{\partial\tau}{\partial r}p_T
\mathrm {cos}\phi\right)\nn\\
&&\times f_{\mathrm {unstable}}\Gamma_{V\rightarrow \mu^+\mu^-}
/\Gamma_V^{\mathrm {tot}}
\nn\\ 
\label{f.o.}
\eea
where $\Gamma_V^{\mathrm {tot}}$ is the total decay width of the 
vector meson $V$.

\section{Dilepton spectra at RHIC and LHC}

\begin{figure}
\centerline{\includegraphics[scale=0.6]{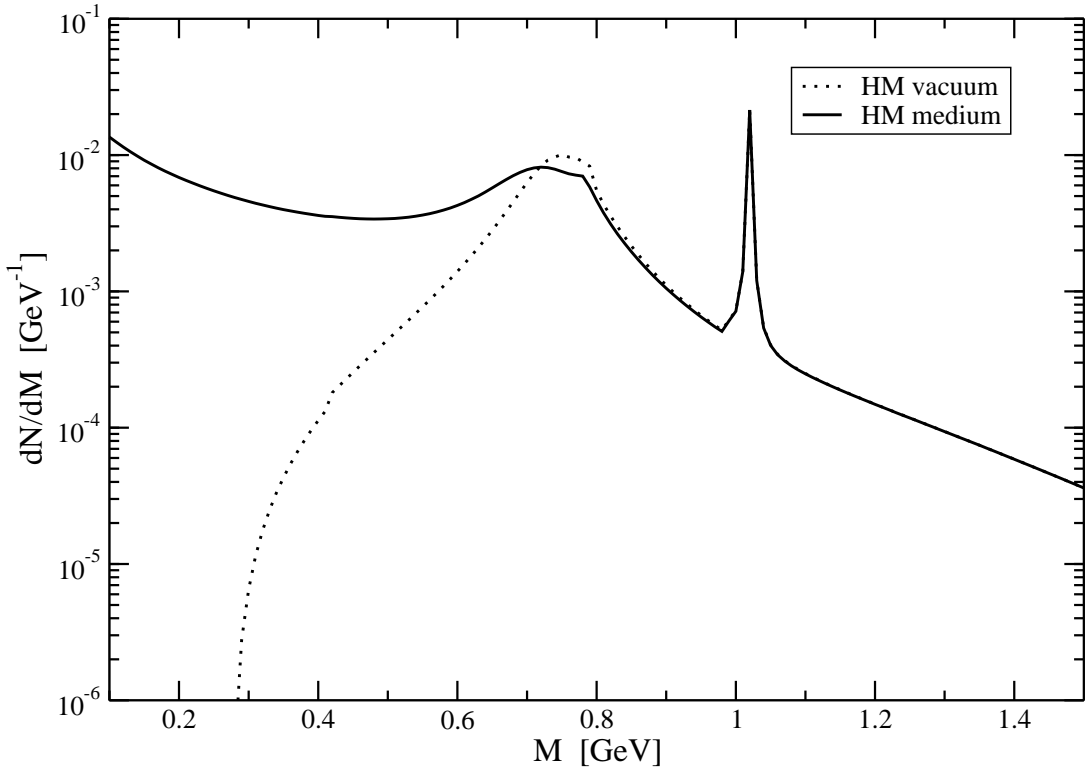}}
\caption{Invariant mass distribution of dileptons from 
hadronic matter (HM) for modified and unmodified  $\rho$ meson.
Scenario (a) has been used for EoS.}
\label{fig:modrho}
\end{figure}

We begin by plotting the space-time integrated invariant mass spectra of 
dileptons. In
Fig.~\ref{fig:modrho} we plot the yield of lepton pairs from the hadronic
matter (HM) containing the effects of chemical non-equilibration both 
in the pole as well as in the continuum part of the spectral function, evaluated with and
without the modified $\rho$ spectral function discussed in Section 2 
for RHIC. 
The enhancement in the region $0.1\leq M\leq 0.7$ GeV is purely a medium effect
and is a contribution from the Landau cut of the $\pi-\omega,h_1,a_1$ loops. 
In simple terms,
the yield in this region results from the scattering of the $\rho$ with 
thermal $\omega$, $h_1$,   $a_1$ and pions~\cite{GhoshEPJC}.
As seen from Eq.~(\ref{eq:impi}) these contributions are weighted by the 
Bose distribution
functions for the pions (minus a much smaller contribution from the
heavy mesons). In contrast, the vacuum spectral function naturally starts
from the $2m_\pi$ threshold coming from the unity in the unitary cut
contribution. 
The (small) kink at 0.42 GeV in this curve is due to the $3m_\pi$
threshold for $\omega$ production. The enhancement in the yield due to 
medium effects is $\sim 20$ for $M$ around 400 MeV. 
\begin{figure}
\centerline{\includegraphics[scale=0.6]{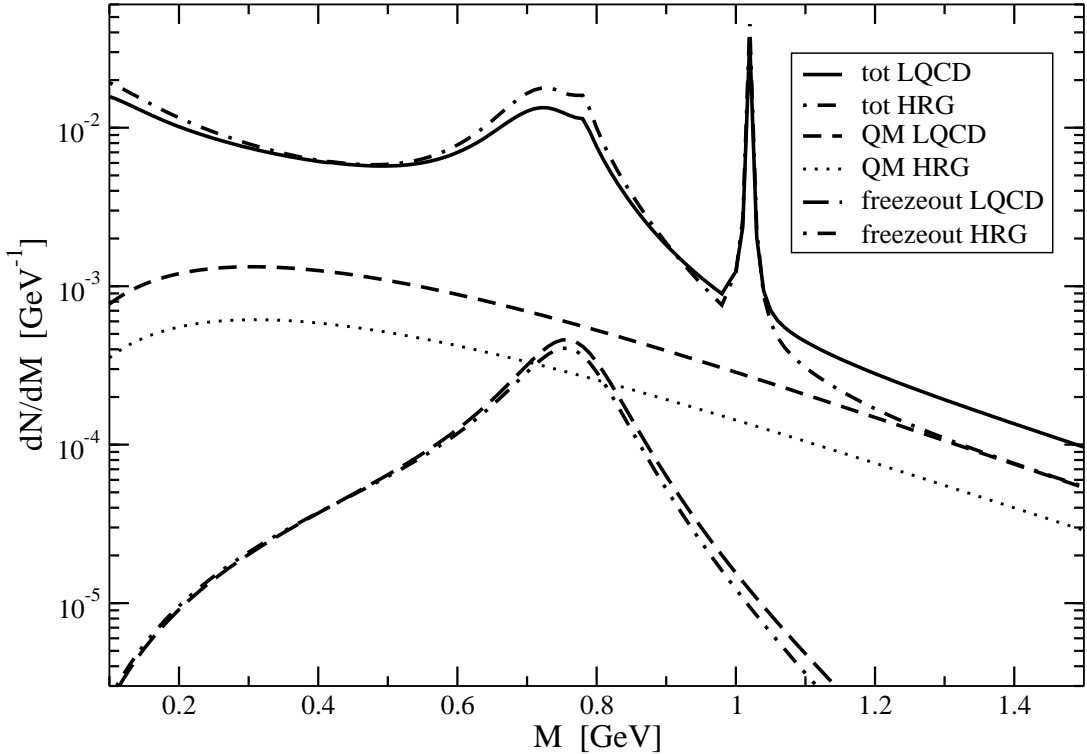}}
\caption{The invariant mass distribution of lepton 
pairs for RHIC initial condition with HRG  and LQCD EoS.}
\label{fig:eosdep}
\end{figure}
Next we show in Fig.~\ref{fig:eosdep} the dependence of the yield from the 
two phases on the EoS. Dilepton radiation from hadronic phase 
outshines the emission from quark matter for $M$ up to $\phi$ mass.
Since we have included the continuum in the $\rho$ and $\omega$ channels 
we ignore the 
four pion annihilation process~\cite{4pi} to avoid double counting.
The contributions from quark matter phase 
dominates over its hadronic counter part for both the EoS
for $M$ beyond $\phi$-peak. This fact may be used to extract 
various properties {\it i.e. average flow, temperature etc.} 
of quark matter and hadronic matter   by 
selecting $M$ windows judiciously.
The dilepton yield from hadronic matter is observed to be larger when the 
HRG EoS is employed in comparison to LQCD. 
This can be understood in terms of the velocity of
sound $v_s^2(=dP/d\epsilon $ evaluated at constant entropy)
 which controls the rate of expansion. 
For EoS of the type (a) $v_s^2\sim 1/3$ in the QGP phase 
which is larger than 
the value of the corresponding quantity for EoS of the type (b).  
Therefore, the rate of expansion in the scenario (b) is
comparatively slower, allowing the QGP to emit lepton pairs for a longer 
time resulting in greater yield for LQCD EoS. 
In contrast, for the  EoS (a), the lower value of $v_s$ 
for the hadronic phase results in a slower cooling and hence a larger yield. 
Also shown for comparison is the yield from the decays of $\rho$ mesons
at the freeze-out for the two types of EoS used. The yield from this source is 
much smaller and we  will not consider it any further.

\begin{figure}
\centerline{\includegraphics[scale=0.6]{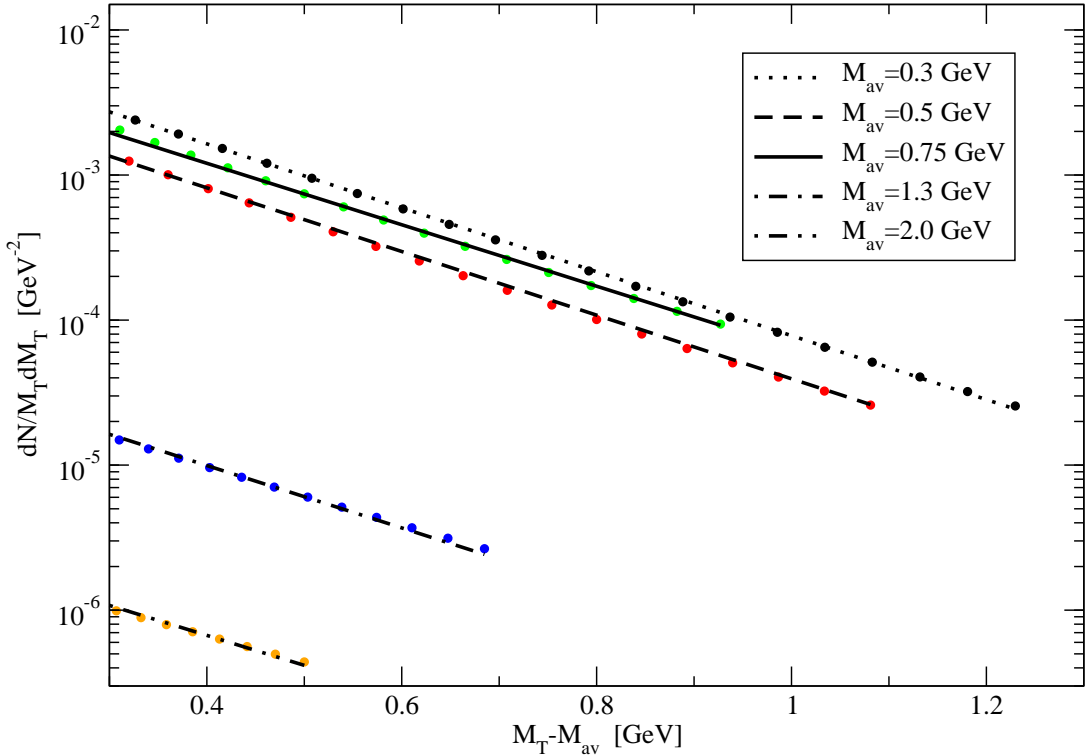}}
\caption{The dilepton yield plotted against $M_T-M_{av}$ for different $M$
windows for RHIC initial conditions.}
\label{fig:mtdist}
\end{figure}

\begin{figure}
\centerline{\includegraphics[scale=0.6]{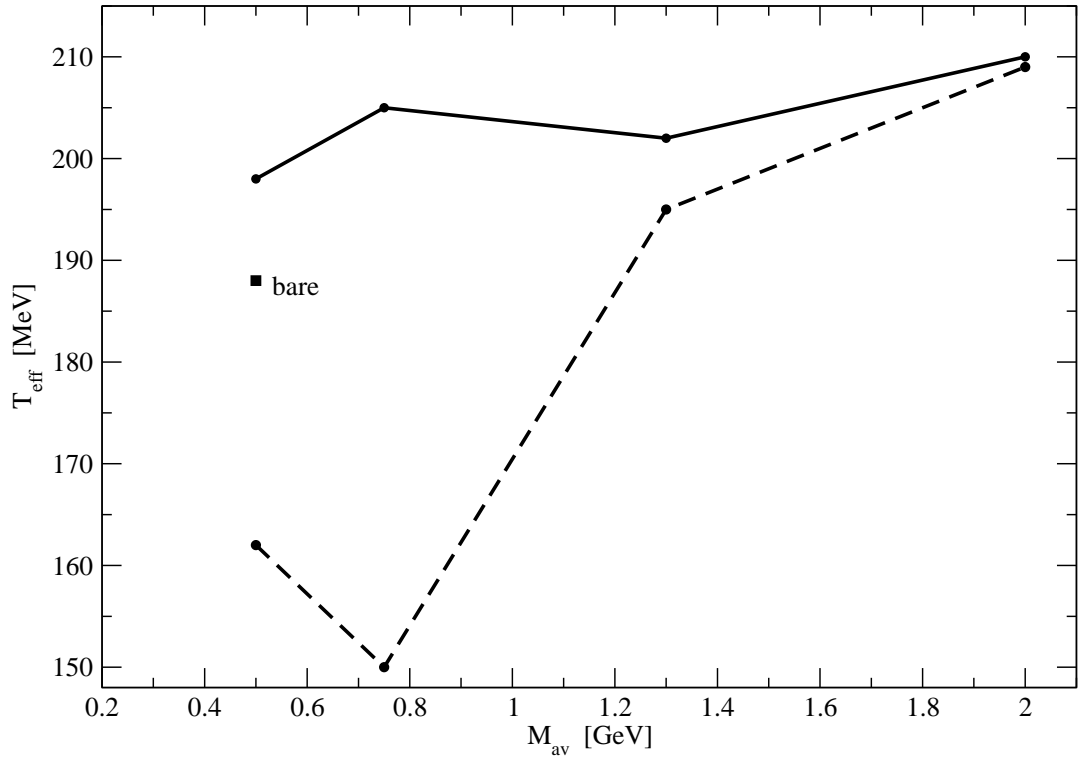}}
\caption{(Colour online) $T_{eff}$ for different values of the $M$-bins for RHIC energy. 
The dashed line is obtained by setting $v_T=0$.}
\label{fig:teff}
\end{figure}

Since the $M$ spectra is invariant under flow
we now turn to the $M_T(=\sqrt{p_T^2+M^2_{\mathrm av}}$)
spectra to study this aspect. Fig.~\ref{fig:mtdist}
shows the $M_T$ spectra of lepton pairs at RHIC energies. 
Here the differential
yield is integrated over small bins of the pair invariant mass 
(from $M_1$ to $M_2$) and plotted
against $M_T-M_{av}$ which is  actually a measure of the 
average kinetic energy (KE) 
of the pair, $M_{av}$ being the average mass ($=[M_1+M_2]/2$) of the bin. 
The average value of $M_T$ for a static system at a temperature 
$T$ is given by $\langle M_T\rangle\sim M+T$. Therefore, the
average KE $\sim T$, is the slope of the $M_T$ distribution. 
Initially, the entire energy of the system formed in HIC is thermal in nature
and with progress of time some part of the thermal energy gets 
converted to the collective (flow) energy.
In other words, during the expansion stage the total energy of the system is shared by the thermal
as well as the collective degrees of freedom.
As a consequence, unlike the invariant mass spectra, the $M_T$ (or $p_T$) 
spectra is heavily influenced by the collective flow  and
the average KE or the inverse slope may be written as 
$T_{\mathrm eff}=T+1/2 M_{\mathrm av} v_T^2$, where $v_T$ is the
average radial flow velocity.
The $M_T$ spectra of dileptons  for various $M$-bins  
have an exponential nature, the inverse
slope providing an effective temperature, $T_{eff}$.
 It is important to mention at this point that for a radially expanding system
the $T_{\mathrm eff}$ has an explicit (linear) $M$ dependence as mentioned
above. However, it has also an implicit $M$ dependence even when 
$v_T=0$ ({\it i.e.} with longitudinal expansion only)
because it is expected that the high (low) $M$ pairs predominantly 
emit from the high (low) temperature or early (late) time zone. For a radially
expanding system
the $M$ dependence of the $T_{\mathrm eff}$ is stronger than for a system which
expands longitudinally only.    

In Fig.~\ref{fig:teff} we have plotted the effective temperature versus
$M_{av}$ for various mass windows of the lepton pairs at RHIC energies,
 evaluated with the
in-medium spectral function of the vector mesons. Also shown by
a filled square is the value of $T_{eff}$ for the vacuum case in the
window $0.4\leq M\leq 0.6$ where there is substantial difference between
the yields in free and medium cases as seen in Fig~\ref{fig:modrho}.
The slope of these curves measure the average temperature and the flow of 
the matter.

Let us try to understand the non-monotonic variation of the
inverse slope with $M_{\mathrm av}$ depicted in Fig.~\ref{fig:teff}.
In Fig.~\ref{fig:eosdep} it is  shown that the 
high $M$ (above $\phi$ peak) pairs originate 
predominantly from the  partonic source and 
the low $M$ (below $\rho$ mass) domain, although 
outshine by the radiation from hadronic source, 
contains non-negligible contributions from quark matter  
{\it i.e.}  the low $M$ region contains
contributions both from the hadronic as well as  QGP phases. 
Now, the collectivity (or flow) in the system 
does not develop fully in the 
QGP because of the small life time of this phase
which means that the radial velocity extracted from
the high $M$ region is small. Here the temperature decreases mainly due to 
longitudinal expansion and 
consequently, the effective slope decreases slowly with
decreasing $M_{\mathrm av}$.  In contrast, the lepton pairs 
with mass around $\rho$-peak dominantly originate from the 
hadronic source 
(which appears in the late stage of the evolving system) 
and are significantly
affected by the flow resulting in higher values of
$v_T$ and hence a higher $T_{\mathrm eff}$.  
At still lower values of $M (<m_\rho$)  the contributions from 
hadronic phase is considerable. Therefore, the $M_T$ spectra in this
domain is affected by the flow significantly  {\it i.e.} the value
of $v_T$ is large but can not be as large as in the $\rho$-peak region  
despite the substantial medium induced
enhancement of the hadronic sources since this domain also contains
contribution from the QGP. 
But for low $M (<m_\rho$) the $T_{\mathrm eff}$ is smaller compared to
the  $\rho$ peak region because of the smaller values $M$. 
The value of $T_{\mathrm eff}$ increases linearly with $M$ up to the $\rho$ peak,
a behaviour typical of inverse slope extracted from the
transverse momentum spectra of hadrons with different masses. 
Thus the values of $T_{\mathrm{eff}}$ for $M$ below and above the
$\rho$-peak are smaller compared to the values around the $\rho$ peak
even in the presence of medium effects,
resulting in the non-monotonic behaviour as displayed in Fig.~\ref{fig:teff},
for $0.5\,<$ M(GeV) $\,< 1.3$. 

\begin{figure}
\centerline{\includegraphics[scale=0.6]{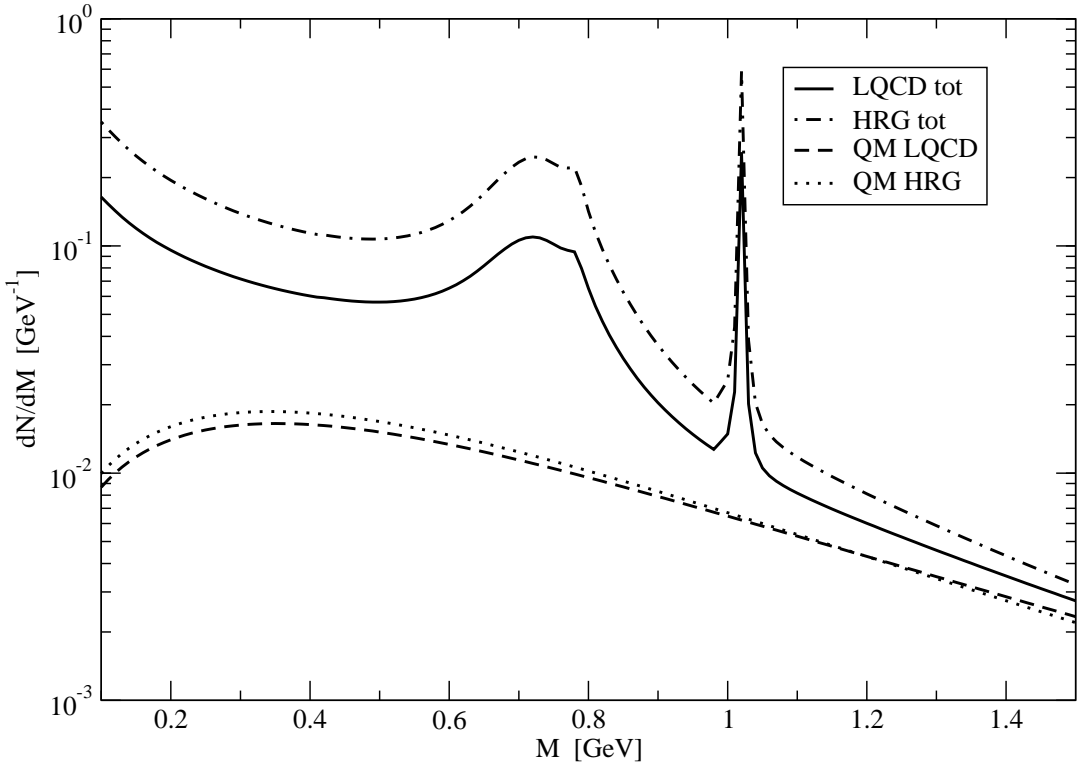}}
\caption{Dilepton yields for HRG EoS and LQCD EoS. 
The initial condition is taken for LHC energy.}
\label{fig:lhc_eosdep}
\end{figure}
\begin{figure}
\centerline{\includegraphics[scale=0.6]{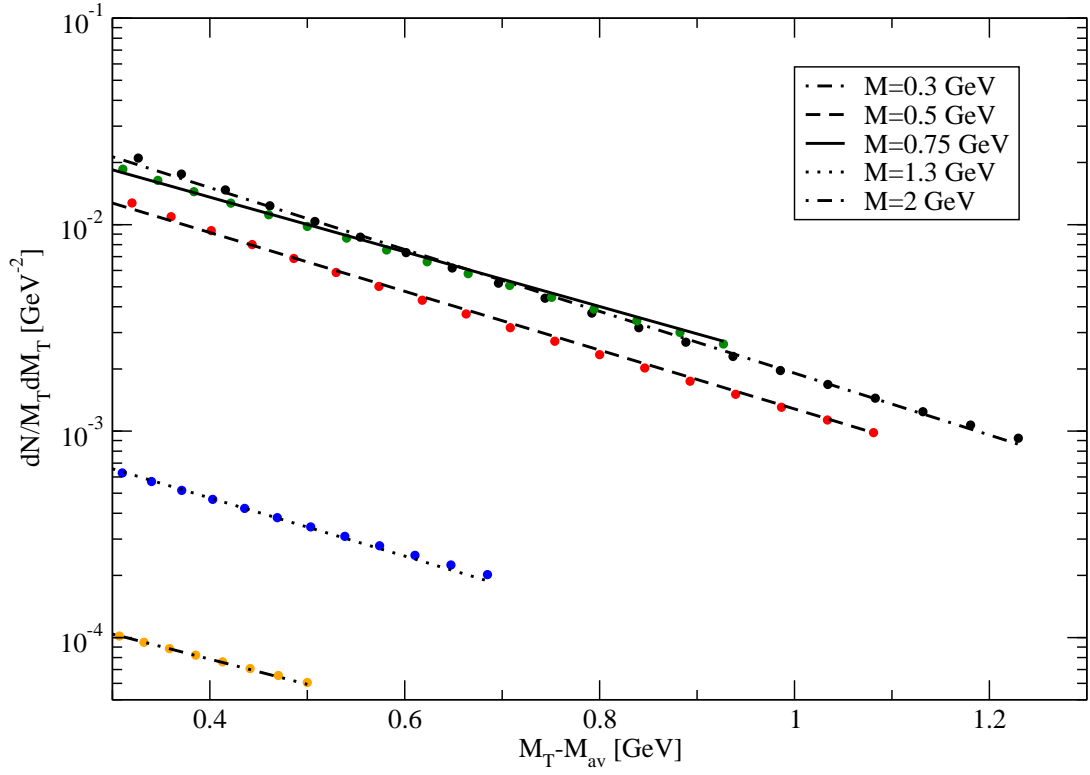}}
\caption{(Colour Online) The dilepton yield plotted against $M_T-M_{av}$ for different $M$
windows for LHC initial condition.}
\label{fig:lhc_mtdist}
\end{figure}
The slope of the $M_T$ spectra is connected with the average collective flow.
It is well known that the average magnitude of
radial flow at the freeze-out surface can be extracted from
the $p_T$ spectra of the hadrons.
However, hadrons being strongly interacting objects
can bring the information of the state of the system
when it is too dilute to support collectivity {\it i.e.}
the parameters of collectivity extracted from the hadronic
spectra are limited to the evolution stage where the
collectivity ceases to exist. These collective parameters have
hardly any information about the interior of the matter.
On the other hand the dileptons are produced and emitted
from all space time points.
Therefore, the value of $v_T$ estimated from the dilepton spectra  will
be lower than the value extracted from the hadronic spectra~\cite{sanja}.
Indeed, the values of $v_T$ estimated from the slopes
of the curve is 0.25 for the $M$ domains $0.5\,<M\,$(GeV)$<0.77$. 
This value is much smaller than the value of $v_T$ 
extracted from the hadronic spectra~\cite{xnu}.  
The dashed line in Fig.~\ref{fig:teff} is obtained 
by setting $v_T=0$. The results indicate that the 
observed (solid line) rise (for $0.5\,<M\,$(GeV)$<0.77$) and  fall 
(for $0.77\,<M\,$(GeV)$<1.3$) 
are due to radial expansion of the system. However,
the rise in large $M$ domain is due to cooling 
of the system due to longitudinal expansion - which
is described as the implicit $M$ dependence of $T_{\mathrm eff}$ above.

 
\begin{figure}
\centerline{\includegraphics[scale=0.6]{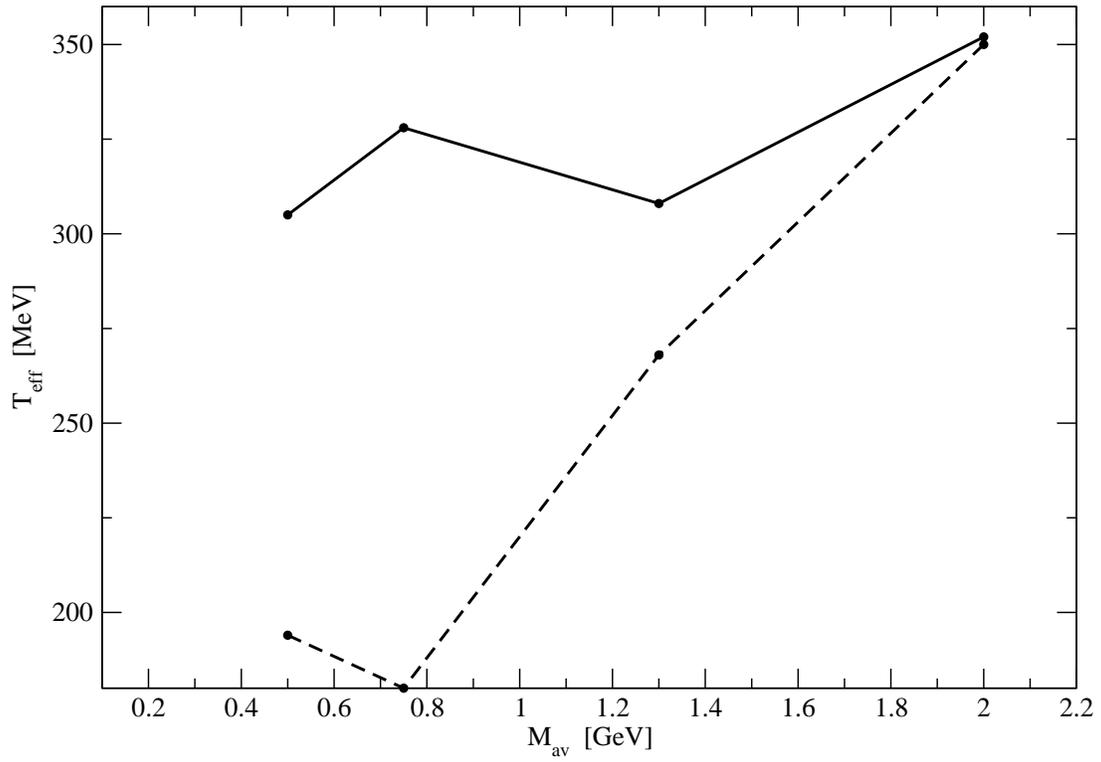}}
\caption{$T_{eff}$ for different values of the $M$-bins for LHC conditions.
The dashed line is obtained by setting $v_T=0$.}
\label{fig:lhc_teff}
\end{figure}

The invariant mass spectra of lepton pairs is
displayed for LHC initial conditions in Fig.~\ref{fig:lhc_eosdep}.
Although, the results are qualitatively similar to RHIC,
quantitatively the yield at LHC is larger.  
This is because of the larger four-volume of the system  to be
realised at  LHC resulting from  a higher value of $T_i$ 
for fixed $T_c$ and $T_F$. Similar kind of 
enhancement is also reflected in the transverse mass distributions of
the lepton pairs at LHC (Fig.~\ref{fig:lhc_mtdist}).

Finally the variation of inverse slope of the $M_T$ distributions
with $M_{\mathrm av}$ for LHC is depicted in Fig.~\ref{fig:lhc_teff}.
The values of $T_{\mathrm eff}$ for various $M$-bins are
larger than RHIC because of the combined effects of large initial
temperature and flow. In fact the value of $v_T$ for 
$0.5<$ M(GeV)$<0.77$ is $\sim 0.52$ compared to 0.25 at RHIC. 
The radial flow in the system is responsible for the 
rise and fall of  $T_{\mathrm eff}$ with $M_{\mathrm av}$ 
(solid line) in the mass region  ($0.5<$ M(GeV)$<1.3$),
for $v_T=0$ (dashed line) a completely different behaviour
is obtained. 
This type of non-monotonic variation of $T_{\mathrm eff}$ 
can not be obtained with a single dilepton source ~\cite{renk}. 
Therefore, such non-monotonic variation of 
the inverse slope deduced from the transverse mass  distribution of 
lepton pairs with average invariant mass is an indication of
the presence of two different phases during the evolution of the 
system. Thus, such variation may be treated as a signal of QGP
formation in heavy ion collisions.  A comment on the thermal emission
rate from the QGP is in order here. We have considered 
the lowest order processes to evaluate the the lepton pair productions
form QGP. However, we have checked that the inclusions of the processes of
order $O(\alpha\alpha_s)$~\cite{braaten,thoma,altherr} changes
the slope parameter, $T_{\mathrm eff}$ at low $M$ by negligible amount. 
It is less than 1\%  for  
$M\sim 0.3$ GeV  and  at higher $M$ it is vanishingly small. 

The other sources of dileptons
{\it e.g.} from the Drell-Yan (DY) mechanism and charm decays
may provide significant ``background'' to the thermal productions
at high mass region ($2\leq M$ (GeV) $\leq 6$ ~\cite{vogt})
which are neglected here because in the present work we focus mainly 
on the low mass regions. Moreover, the contributions from the 
DY process and charm decays from proton+proton (pp) collisions  may 
be used to estimate the similar contributions from heavy ion 
collisions at the same colliding energy by appropriately scaling 
pp data by the effective number  of nucleon+nucleon collisions in nuclear
interaction.  The contributions from the $\rho$  at the freeze-out
surface has been evaluated and it is found to be small. 

\section{Conclusion}

In this work we have attempted to bring out distinguishing features
stemming from many body effects in the lepton pair yield from relativistic heavy ion collisions.
On the microscopic side we have used a
$\rho$ spectral function evaluated at finite 
temperature in the framework of real time formalism of thermal field theory
using interaction vertices from chiral perturbation theory. 
The effect of baryons have also been included in the empirical 
approach of Ref.~\cite{Eletsky} using resonance
dominance in the forward scattering amplitude. 
A significant enhancement of the dilepton yield in the region below the 
nominal $\rho$ peak has been obtained from the combined effect of mesons and 
baryons. The space-time evolution of the dilepton emission rate 
using relativistic hydrodynamics with
chemical freeze-out reflects this enhancement in the invariant mass spectra.  

On the macroscopic side, it is argued that the non-monotonic variation of 
the inverse slope deduced from the transverse mass distribution of 
lepton pairs for various values of the average invariant mass is an 
indication of the presence of two different phases during the evolution of the 
system. Thus, such a variation may be treated as a signal of QGP
formation in heavy ion collisions.  


{\bf Acknowledgement:}
JA's research is partially supported by DAE-BRNS project No.  
2005/21/5- BRNS/2455.

\end{document}